\documentclass[11pt]{article}

\usepackage[utf8]{inputenc}
\usepackage[T1]{fontenc}
\usepackage{lmodern}
\usepackage[a4paper,margin=1in]{geometry}
\usepackage{amsmath,amssymb}
\usepackage{booktabs}
\usepackage{enumitem}
\usepackage{hyperref}
\usepackage{url}
\usepackage{graphicx}
\usepackage{xcolor}
\usepackage[numbers]{natbib}
\usepackage{microtype}

\hypersetup{
  colorlinks=true,
  linkcolor=blue,
  citecolor=blue,
  urlcolor=blue,
  pdftitle={ESAA-Security: An Event-Sourced, Verifiable Architecture for Agent-Assisted Security Audits of AI-Generated Code},
  pdfauthor={Elzo Brito dos Santos Filho}
}

\title{ESAA-Security: An Event-Sourced, Verifiable Architecture for Agent-Assisted Security Audits of AI-Generated Code}
\author{Elzo Brito dos Santos Filho\\elzo.santos@cps.sp.gov.br}
\date{}

\begin{document}
\maketitle

\begin{abstract}
AI-assisted software generation has increased development speed, but it has also amplified a persistent engineering problem: systems that are functionally correct may still be structurally insecure. In practice, prompt-based security review with large language models often suffers from uneven coverage, weak reproducibility, unsupported findings, and the absence of an immutable audit trail. The ESAA architecture addresses a related governance problem in agentic software engineering by separating heuristic agent cognition from deterministic state mutation through append-only events, constrained outputs, and replay-based verification \citep{esaa2026,eventSourcing2005,cqrs2010}.

This paper presents ESAA-Security, a domain-specific specialization of ESAA for agent-assisted security auditing of software repositories, with particular emphasis on AI-generated or AI-modified code. ESAA-Security structures auditing as a governed execution pipeline with four phases---reconnaissance, domain audit execution, risk classification, and final reporting---and operationalizes the workflow into 26 tasks, 16 security domains, and 95 executable checks \citep{esaaSecurityRepo}. The framework produces structured check results, vulnerability inventories, severity classifications, risk matrices, remediation guidance, executive summaries, and a final markdown/JSON audit report.

The central idea is that security review should not be modeled as a free-form conversation with an LLM, but as an evidence-oriented audit process governed by contracts and events. In ESAA-Security, agents emit structured intentions under constrained protocols; the orchestrator validates them, persists accepted outputs to an append-only log, reprojects derived views, and verifies consistency through replay and hashing. The result is a traceable, reproducible, and risk-oriented audit architecture whose final report is auditable by construction.
\end{abstract}

\section{Introduction}

LLM-based software engineering is moving from isolated assistance toward agentic workflows that must preserve consistency across files, tasks, and long execution horizons. Recent work has shown that language models can reason and act through tools, navigate repositories, and participate in multi-agent execution settings \citep{react2023,swebench2024,autogen2023,metagpt2023}. However, this capability shift also exposes a governance problem: once agents operate over long horizons, mutable state, incomplete context, and unverifiable intermediate steps become operational risks rather than mere prompting inconveniences. The ESAA architecture responds to this problem by treating the event log, not the mutable repository snapshot, as the source of truth, and by deriving project state through deterministic projection, replay, and verification \citep{esaa2026,eventSourcing2005,cqrs2010}.

That shift is particularly relevant in security auditing. In AI-generated software and so-called vibe coding contexts, development speed has increased, but so has the incidence of systems that appear functionally acceptable while remaining weak in authentication, authorization, input validation, secret handling, session control, dependency hygiene, infrastructure hardening, and AI-specific risk handling \citep{owaspTop10,asvs,esaaSecurityRepo}. The problem is not merely that models can miss vulnerabilities; it is that ad hoc prompt-based review usually leaves scope implicit, findings weakly structured, severity inconsistently assigned, and final conclusions difficult to reproduce or audit \citep{esaaSecurityRepo}.

ESAA-Security is proposed as a domain-specific answer to that problem. Rather than asking an LLM to ``review this repository for vulnerabilities'' in open prose, ESAA-Security models security review as a governed execution pipeline. The framework inherits ESAA's event-sourced governance kernel and specializes it for security analysis by defining dedicated phases, task semantics, playbooks, contracts, and output products. In this architecture, findings do not become part of the audit merely because a model wrote persuasive prose. They become part of the audit only after being emitted in structured form, validated under contract, appended to the event log, incorporated into projected views, and preserved under replay-verifiable integrity \citep{esaa2026,esaaSecurityRepo}.

The core claim of this paper is therefore architectural: agent-assisted security review should be treated as a governed execution problem rather than merely a prompting problem. The relevant question is not only whether vulnerabilities are mentioned, but whether the resulting security assessment is explicit in scope, structured in evidence, reproducible under replay, and traceable from check-level findings to final conclusions. ESAA-Security is designed precisely around that shift.

The contributions of this paper are fourfold:
\begin{enumerate}[leftmargin=*,itemsep=0.3em]
  \item We present ESAA-Security as a specialization of the ESAA architecture for evidence-based security auditing of software repositories, especially those produced or heavily modified with AI assistance.
  \item We define a governed audit pipeline that combines constrained agent outputs, append-only event persistence, deterministic reprojection, and replay-based verification.
  \item We operationalize security review into four phases, 26 tasks, 16 domains, and 95 executable checks, producing structured outputs from check-level findings to a final risk-oriented report.
  \item We define an evaluation framing in which the relevant outcome is not only vulnerability enumeration, but also traceability, reproducibility, coverage explicitness, artifact completeness, and remediation usefulness.
\end{enumerate}

\section{Background and Related Work}

\subsection{Agentic Software Engineering and the Governance Gap}

Recent progress in LLM-based software engineering has shifted the field from isolated conversational assistance toward agentic workflows capable of planning, tool use, repository navigation, and multi-step execution \citep{react2023,swebench2024,autogen2023,metagpt2023}. This transition increases capability, but it also exposes a structural problem: software work requires consistency across artifacts, tasks, and state transitions over long horizons, whereas language models remain probabilistic generators operating under context constraints. As argued in the ESAA architecture paper, the key difficulty is not merely improving prompts, but establishing a governance kernel that can preserve accountability, reversibility, and state coherence under repeated autonomous actions \citep{esaa2026}.

The problem becomes more severe under long-context conditions. Even when large context windows are available, models do not reliably use all relevant information with equal fidelity; performance often degrades when crucial information appears in the middle of long inputs \citep{lostMiddle2024}. For agentic software workflows, this means that instructions, contracts, and earlier evidence can be silently deprioritized. ESAA's notion of purified views and staged execution responds directly to this risk by minimizing uncontrolled context accumulation and shifting consistency enforcement from model memory to orchestrator logic \citep{esaa2026}.

\subsection{Structured Outputs, Deterministic Validation, and State Traceability}

A second relevant line of work concerns structured generation and output validation. Modern LLM systems increasingly rely on structured or schema-constrained outputs to reduce parsing ambiguity and make model outputs machine-actionable \citep{jsonSchemaOutputs2024,grammarDecoding2024,structuredOutputs2024}. Grammar-constrained decoding and schema-aware generation reduce syntax errors and improve downstream reliability, but they do not by themselves define how state should evolve across multi-step workflows. ESAA extends this line by requiring agents to emit structured intentions that are validated against contracts and then applied by a deterministic orchestrator, rather than allowing the model to mutate system state directly \citep{esaa2026}.

At the architectural level, ESAA is grounded in Event Sourcing and CQRS \citep{eventSourcing2005,cqrs2010}. Event Sourcing treats the append-only event log as the source of truth, while current state is reconstructed as a projection of those events. CQRS complements this by separating write-oriented effect application from read-oriented materialized views. ESAA transposes these principles into agentic software engineering: the agent emits intentions, the orchestrator validates them, accepted events are persisted, and derived views are rebuilt and verified through replay. ESAA-Security directly inherits this model and applies it to security auditing.

\subsection{Security Review, Application Security Standards, and AI-Specific Risk}

The substantive security coverage of ESAA-Security is aligned with widely used application security references such as the OWASP Top 10 and the OWASP Application Security Verification Standard (ASVS) \citep{owaspTop10,asvs}. These references are important because they anchor the framework's domain coverage in recognized risk families and verification categories rather than in purely ad hoc prompt phrasing. ESAA-Security extends this alignment with an explicit AI/LLM security domain to account for risks that conventional AppSec frameworks do not always foreground in repository review \citep{esaaSecurityRepo}.

The appropriate way to position ESAA-Security is therefore not as a replacement for ESAA, nor as merely another checklist. It is a domain-specific specialization of the ESAA governance kernel for evidence-based security auditing of software repositories. ESAA contributes the general mechanism: append-only event logs, deterministic orchestration, replay verification, and contract-restricted outputs \citep{esaa2026}. ESAA-Security contributes the domain specialization: security-focused task semantics, encoded audit coverage, structured audit artifacts, and risk-oriented outputs \citep{esaaSecurityRepo}.

\section{ESAA-Security Architecture}

ESAA-Security is designed as a domain-specific specialization of the ESAA governance kernel for security auditing of software repositories, especially repositories produced or heavily modified with AI assistance. The design objective is to convert security review from an open-ended prompt interaction into a governed workflow whose state is explicit, replayable, and auditable. The specialization preserves ESAA's trace-first model and redefines the artifact space around security evidence and risk-oriented reporting \citep{esaa2026,esaaSecurityRepo}.

At a high level, ESAA-Security is organized around five cooperating layers: the audit roadmap, the security playbooks, the agent and orchestrator contracts, the append-only event store, and the projected read-models. The repository structure makes those layers explicit through the roadmap, contracts, runtime policies, playbooks, and phase-specific \texttt{reports/} hierarchy \citep{esaaSecurityRepo}. The workflow is divided into four audit phases:
\begin{enumerate}[leftmargin=*,itemsep=0.3em]
  \item \textbf{Reconnaissance}: identify technology stack, architecture, data flows, and attack surfaces.
  \item \textbf{Domain audit execution}: run playbook-driven audits over the repository, one task per security domain.
  \item \textbf{Risk classification}: consolidate findings into vulnerability inventories, severity assignments, and risk matrices.
  \item \textbf{Final reporting}: generate technical remediations, best-practice guidance, executive summaries, and final audit reports.
\end{enumerate}

Coverage is explicit rather than implicit. ESAA-Security defines 16 domains---including secrets and configuration, authentication, authorization, input validation, dependencies and supply chain, API security, file upload, cryptography, AI/LLM security, and DevSecOps---covering a total of 95 executable checks across 26 tasks in four phases \citep{esaaSecurityRepo,owaspTop10,asvs}. This encoded coverage is one of the architecture's main differentiators from prompt-only review, because it constrains what the system is expected to inspect instead of relying on ad hoc recall by the model.

The role of the orchestrator remains central. Agents do not directly mutate the audit state. They emit structured intentions under explicit contracts, and those intentions are validated before any effect is persisted. Accepted outputs are appended to the event store, projected into current-state views, and verified through replay and hashing. The final report is therefore not a free-form narrative authored from scratch, but the terminal product of a governed sequence of admissible state transitions \citep{esaa2026,esaaSecurityRepo}.

\section{Execution Protocol and Audit Invariants}

The execution protocol of ESAA-Security is designed to guarantee that audit progress is not inferred from model prose but from a validated sequence of admissible state transitions. The current protocol generation uses a two-step-per-task discipline: each task must be claimed before work is performed, and only later completed with verification evidence and admissible artifact writes \citep{esaa2026,esaaSecurityRepo}. This serialized structure makes task ownership, sequencing, and verification explicit.

The first invariant is \emph{claim-before-work}. When a task is in \texttt{todo}, the permitted action is \texttt{claim}; substantive work and file updates are forbidden at this stage. The second invariant is \emph{complete-after-work}. Once the task is in \texttt{in\_progress} and assigned to the same actor, the agent may emit \texttt{complete}, accompanied by verification checks and bounded file updates. The third invariant is \emph{prior-status consistency}: the agent must restate the status it believes it is operating under, which allows the orchestrator to reject stale-context or out-of-phase execution attempts before any append occurs.

A fourth invariant is \emph{lock ownership}. Completion is valid only when the current actor owns the task. A fifth invariant is \emph{boundary discipline}: artifact writes are permitted only under \texttt{complete} and only within the admissible boundary of the current task kind. A sixth invariant is \emph{done immutability}: once a task is terminal, it must not be silently reopened; corrections must route through explicit issue or hotfix flows. Together, these invariants make the audit trail a governed state machine rather than an uncontrolled sequence of textual claims.

The validation model is explicitly fail-closed. Invalid transitions, schema violations, boundary violations, status mismatches, or attempts to collapse multiple actions into a single emission are rejected before they affect state. This is essential to the framework's trust model: the event log is authoritative only if invalid outputs cannot silently contaminate it. Replay-based verification then acts as a final integrity layer, ensuring that projected audit state can be reconstructed from admitted events alone \citep{esaa2026,eventSourcing2005,cqrs2010}.

\section{Audit Outputs and Risk-Oriented Reporting}

The output model of ESAA-Security is phase-typed and cumulative. Phase 2 produces domain narratives and structured check results. Phase 3 produces a vulnerability inventory, classified vulnerabilities, and a risk matrix. Phase 4 produces technical remediations, best-practice guidance, and an executive summary. The terminal outputs are a human-readable final report and a structured JSON report \citep{esaaSecurityRepo}. This partition matters because the final report is not the primary object produced by the audit; it is the final projection of an evidence pipeline.

At the lowest output layer, check-level findings are structured evidence objects rather than free-form narrative claims. Each finding is expected to bind together a check identifier, status, severity, code or configuration evidence, technical explanation, and remediation guidance. This is important both epistemically and operationally: it constrains what counts as a valid audit observation and makes downstream consolidation possible.

The next layer is the vulnerability inventory, where domain-local findings are transformed into system-level security state. From there, ESAA-Security performs severity classification using categories such as \texttt{CRITICAL}, \texttt{HIGH}, \texttt{MEDIUM}, \texttt{LOW}, and \texttt{INFO}, along with CIA-oriented impact reasoning. A risk matrix then orders findings by severity, impact, and remediation relevance. This matrix becomes the pivot from analysis to action.

The recommendation layer contains two distinct outputs. Technical remediations provide concrete engineering guidance for high-priority issues. Best-practice outputs organize recurring lessons and hardening guidance by domain. The executive summary compresses the audit into a decision-oriented artifact, including a 0--100 security score and top-priority risks. Finally, the full report compiles all of these layers into a structured narrative whose sections remain traceable to check-level findings and intermediate artifacts \citep{esaaSecurityRepo}.

This reporting cascade is one of the strongest distinctions between ESAA-Security and prompt-only review. In an ad hoc review, the model's narrative is often treated as the audit product itself. In ESAA-Security, narrative is downstream of evidence and state. The final report is authoritative only insofar as it is the projection of an admitted event sequence and its dependent artifacts.

\section{Evaluation Design and Research Questions}

The right evaluation objective for ESAA-Security is not simply to ask whether a model can name plausible vulnerabilities. It is to determine whether a security audit can be executed as a governed, replay-verifiable process that produces structured findings, risk-oriented outputs, and final reports traceable to immutable events. This framing follows directly from ESAA's use of the orchestrator run as the primary empirical unit and from ESAA-Security's insistence that report integrity depends on verification status \citep{esaa2026,esaaSecurityRepo}.

Three research questions follow naturally:
\begin{enumerate}[leftmargin=*,itemsep=0.3em]
  \item \textbf{RQ1}: Can an event-sourced execution model make agent-assisted security audits replay-verifiable and traceable at the level of findings, classifications, and final reports?
  \item \textbf{RQ2}: Can security review be operationalized as a structured audit process with explicit domain coverage, check-level evidence, and risk-oriented outputs rather than as a free-form prompting activity?
  \item \textbf{RQ3}: Does ESAA-Security produce artifacts that are useful for prioritization and remediation in AI-generated software, including severity classification, risk matrix generation, technical fixes, and executive reporting?
\end{enumerate}

The primary unit of analysis should remain the orchestrator run, consistent with ESAA. However, ESAA-Security also requires a second analytical layer: the artifact chain produced within a run. For each case, evaluation should therefore inspect not only run termination and verification metadata, but also whether the expected chain of Phase 2 results, Phase 3 inventories and matrices, Phase 4 remediation/report artifacts, and final outputs was produced and remained internally consistent with the roadmap dependencies.

Evaluation should cover at least five dimensions: protocol compliance, replay-verifiable state integrity, coverage completeness, artifact completeness, and risk-report usefulness. This design is better aligned with the framework's claims than raw finding count alone. A system that produces many speculative findings without traceability or structured remediation is not necessarily stronger than one that produces fewer findings but yields reproducible, evidence-backed, and decision-oriented outputs.

\section{Initial Validation Plan and Case Study Protocol}

A defensible initial validation should follow the empirical logic of the ESAA paper and use at least two case studies of different scale and complexity \citep{esaa2026}. One case should be a small repository that permits close manual inspection of every artifact. The second should be a medium repository that exercises more domains, dependency paths, and reporting transitions. At least one repository should be substantially AI-generated or AI-modified, since that is the motivating context of the specialization.

Repository selection should satisfy four criteria. First, the repository must expose enough technical surface for reconnaissance to operate meaningfully. Second, it should enable nontrivial execution of multiple security domains. Third, it should be auditable under the declared input contract, ideally with at least repository path and some optional infrastructure or documentation artifacts. Fourth, it should support downstream usefulness analysis by containing enough implementation detail for severity, remediation, and executive reporting to be meaningful.

Each run should begin from initialized audit state and preserve the documented runtime configuration. The execution procedure should follow the canonical cycle: parse event store, project current state, select next eligible task, dispatch the appropriate agent with purified context, validate output, append accepted events, rebuild views, verify integrity through replay and hashing, and continue until all tasks are either completed or explicitly blocked. Success should be defined operationally: verified final state, no unrecoverable protocol violations, and a complete artifact chain for the reached phases.

A natural first baseline is prompt-only review: ask an LLM to review the same repository for vulnerabilities and compare the outputs with ESAA-Security along dimensions such as coverage explicitness, evidence structure, replayability, and report completeness. A second baseline is checklist-only review, where the same domains/checks are followed without the ESAA governance layer. The goal is not necessarily to prove that ESAA-Security always finds more vulnerabilities, but to show that its outputs are more auditable, more structurally complete, and more useful for downstream remediation and reporting.

\section{Discussion, Limitations, and Threats to Validity}

The main implication of ESAA-Security is that the quality of an AI-assisted security audit should be judged not only by how many plausible vulnerabilities are named, but by whether the audit can be reconstructed, checked, and trusted as a governed process. This is a direct consequence of the ESAA trace-first model and of ESAA-Security's requirement that the final report remain a faithful projection of the audit trail \citep{esaa2026,esaaSecurityRepo}.

At the same time, the framework's claims must remain bounded. ESAA-Security does not prove the absence of vulnerabilities, does not replace penetration testing, and should not be interpreted as a certification mechanism. The strongest defensible claim is therefore architectural and methodological: ESAA-Security provides a governed, traceable, and replay-verifiable way to execute security audits in AI-accelerated development environments. That framing is consistent with the project's own positioning around OWASP/ASVS-aligned internal audit maturity rather than full external assurance \citep{owaspTop10,asvs,esaaSecurityRepo}.

One important limitation is dependence on playbook quality. Because the substantive security logic is encoded in playbooks and checks, weak playbooks can yield well-governed but substantively weak audits. Another limitation is dependence on repository context: if infrastructure details, deployment settings, or operational context are absent, the audit may remain protocol-valid while still being context-poor. A third limitation is semantic model error: structured outputs reduce the structure gap, but they do not eliminate misinterpretation of framework conventions, compensating controls, or exploitability. ESAA-Security's guardrails reduce that risk, but they cannot remove it entirely.

There is also a maturity limitation in the current specialization. The ESAA-Security repository is already organized around \texttt{reports/phase1} through \texttt{reports/final}, but inherited generic ESAA profiles may still reflect more implementation-oriented boundaries. This does not weaken the conceptual contribution, but it does matter for implementation maturity and full contract canonization.

A threat to internal validity is that early case studies may confound protocol success with substantive audit quality. A threat to external validity is repository diversity: results from a small number of well-structured repositories may not generalize to large monorepos or poorly documented production systems. A threat to construct validity is evaluation focus: if only finding count is measured, noisier baselines may appear strong; if only protocol cleanliness is measured, the framework may appear strong without demonstrating substantive security value. The correct construct is therefore compound: governed audit quality should include traceability, reproducibility, coverage explicitness, artifact completeness, and remediation usefulness.

\section{Code Availability}

An open-source implementation of ESAA-Security is publicly available at the project repository \citep{esaaSecurityRepo}. The repository includes the contracts, playbooks, roadmap, runtime policies, and reporting artifacts associated with the framework. For formal software citation and long-term archival reproducibility, the recommended practice is to cite a versioned release of the software and, when available, an archived DOI-backed snapshot \citep{force11SoftwareCitation,githubCitationFile,zenodoGithub}.

\section{Conclusion}

This paper presented ESAA-Security as a domain-specific specialization of ESAA for agent-assisted security auditing of software repositories, especially repositories produced or substantially modified with AI assistance. ESAA established the underlying governance kernel through append-only event logs, constrained structured intentions, deterministic orchestration, and replay-verifiable projection. ESAA-Security extends that kernel into the security domain by redefining the workflow around audit phases, security playbooks, structured findings, risk-oriented artifacts, and final reporting products \citep{esaa2026,esaaSecurityRepo}.

The specialization contributes a concrete operational model. Security auditing is organized into four phases, 26 tasks, 16 security domains, and 95 executable checks \citep{esaaSecurityRepo}. Instead of allowing findings to emerge as unconstrained prose, the framework treats them as structured evidence objects and routes them through consolidation, classification, matrix generation, executive summarization, and final report compilation. In doing so, it changes the unit of trust from free-form model opinion to protocol-valid, replay-verifiable audit state \citep{esaa2026,esaaSecurityRepo}.

The strongest contribution of ESAA-Security is therefore not another vulnerability checklist, but a different conception of what an AI-assisted security audit should be: explicit in scope, structured in evidence, constrained in execution, and verifiable in final state. That is a timely and technically meaningful shift for the emerging reality of AI-generated software. Future work should extend the framework empirically through broader case studies and practically through the roadmap's own next steps, including CVSS integration, hybrid DAST, multi-repository auditing, compliance mapping, and temporal comparison across audit runs \citep{esaaSecurityRepo}.


\begin{thebibliography}{99}

\bibitem[Brito dos Santos Filho(2026)]{esaa2026}
Brito dos Santos Filho, E. (2026).
\newblock {ESAA: Event Sourcing for Autonomous Agents in LLM-Based Software Engineering}.
\newblock arXiv preprint arXiv:2602.23193.
\newblock \url{https://arxiv.org/abs/2602.23193}.

\bibitem[Brito(2026)]{esaaSecurityRepo}
Brito, E. (2026).
\newblock {ESAA-Security}.
\newblock GitHub repository.
\newblock \url{https://github.com/elzobrito/ESAA-Security}.

\bibitem[Fowler(2005)]{eventSourcing2005}
Fowler, M. (2005).
\newblock {Event Sourcing}.
\newblock Architectural essay.
\newblock \url{https://www.martinfowler.com/eaaDev/EventSourcing.html}.

\bibitem[Fowler(2011)]{cqrs2010}
Fowler, M. (2011).
\newblock {CQRS}.
\newblock Architectural essay.
\newblock \url{https://martinfowler.com/bliki/CQRS.html}.

\bibitem[Yao et~al.(2023)]{react2023}
Yao, S., Zhao, J., Yu, D., Du, N., Shafran, I., Narasimhan, K.~R., and Cao, Y. (2023).
\newblock {ReAct: Synergizing Reasoning and Acting in Language Models}.
\newblock In \emph{The Eleventh International Conference on Learning Representations (ICLR)}.
\newblock \url{https://openreview.net/forum?id=WE_vluYUL-X}.

\bibitem[Wu et~al.(2024)]{autogen2023}
Wu, Q., Bansal, G., Zhang, J., Wu, Y., Li, B., Zhu, E., Jiang, L., Zhang, X., Zhang, S., Liu, J., Awadallah, A.~H., White, R.~W., Burger, D., and Wang, C. (2024).
\newblock {AutoGen: Enabling Next-Gen LLM Applications via Multi-Agent Conversation}.
\newblock In \emph{First Conference on Language Modeling (COLM)}.
\newblock \url{https://www.microsoft.com/en-us/research/publication/autogen-enabling-next-gen-llm-applications-via-multi-agent-conversation-framework/}.

\bibitem[Hong et~al.(2024)]{metagpt2023}
Hong, S., Zhuge, M., Chen, J., Zheng, X., Cheng, Y., Wang, J., Zhang, C., Wang, Z., Yau, S.~K.~S., Lin, Z., Zhou, L., Ran, C., Xiao, L., Wu, C., and Schmidhuber, J. (2024).
\newblock {MetaGPT: Meta Programming for A Multi-Agent Collaborative Framework}.
\newblock In \emph{The Twelfth International Conference on Learning Representations (ICLR)}.
\newblock \url{https://openreview.net/forum?id=VtmBAGCN7o}.

\bibitem[Liu et~al.(2024)]{lostMiddle2024}
Liu, N.~F., Lin, K., Hewitt, J., Paranjape, A., Bevilacqua, M., Petroni, F., and Liang, P. (2024).
\newblock {Lost in the Middle: How Language Models Use Long Contexts}.
\newblock \emph{Transactions of the Association for Computational Linguistics}, 12:157--173.
\newblock \url{https://aclanthology.org/2024.tacl-1.9/}.

\bibitem[Pokrass(2024)]{jsonSchemaOutputs2024}
Pokrass, M. (2024).
\newblock {Introducing Structured Outputs in the API}.
\newblock OpenAI technical article.
\newblock \url{https://openai.com/index/introducing-structured-outputs-in-the-api/}.

\bibitem[Geng et~al.(2023a)]{grammarDecoding2024}
Geng, S., Josifoski, M., Peyrard, M., and West, R. (2023a).
\newblock {Grammar-Constrained Decoding for Structured NLP Tasks without Finetuning}.
\newblock In \emph{Proceedings of the 2023 Conference on Empirical Methods in Natural Language Processing}, pages 10932--10952.
\newblock \url{https://aclanthology.org/2023.emnlp-main.674/}.

\bibitem[Geng et~al.(2025)]{structuredOutputs2024}
Geng, S., Cooper, H., Moskal, M., Jenkins, S., Berman, J., Ranchin, N., West, R., Horvitz, E., and Nori, H. (2025).
\newblock {Generating Structured Outputs from Language Models: Benchmark and Studies}.
\newblock arXiv preprint arXiv:2501.10868.
\newblock \url{https://arxiv.org/abs/2501.10868}.

\bibitem[Jimenez et~al.(2024)]{swebench2024}
Jimenez, C.~E., Yang, J., Wettig, A., Yao, S., Pei, K., Press, O., and Narasimhan, K.~R. (2024).
\newblock {{SWE}-bench: Can Language Models Resolve Real-World {GitHub} Issues?}
\newblock In \emph{The Twelfth International Conference on Learning Representations (ICLR)}.
\newblock \url{https://openreview.net/forum?id=VTF8yNQM66}.

\bibitem[OpenAI(2024)]{swebenchVerified2024}
OpenAI (2024).
\newblock {Introducing {SWE}-bench Verified}.
\newblock Technical report / milestone post.
\newblock \url{https://openai.com/index/introducing-swe-bench-verified/}.

\bibitem[OWASP Foundation(2021)]{owaspTop10}
OWASP Foundation (2021).
\newblock {{OWASP} Top 10:2021}.
\newblock Official project documentation.
\newblock \url{https://owasp.org/Top10/2021/}.

\bibitem[OWASP Foundation(2025)]{asvs}
OWASP Foundation (2025).
\newblock {Application Security Verification Standard (ASVS) Version 5.0.0}.
\newblock Official standard.
\newblock \url{https://owasp.org/www-project-application-security-verification-standard/}.

\bibitem[Smith et~al.(2016)]{force11SoftwareCitation}
Smith, A.~M., Katz, D.~S., Niemeyer, K.~E., and FORCE11 Software Citation Working Group (2016).
\newblock {Software Citation Principles}.
\newblock \emph{PeerJ Computer Science}, 2:e86.
\newblock \url{https://peerj.com/articles/cs-86/}.

\bibitem[GitHub(2026)]{githubCitationFile}
GitHub (2026).
\newblock {About CITATION files}.
\newblock Official documentation.
\newblock \url{https://docs.github.com/repositories/managing-your-repositorys-settings-and-features/customizing-your-repository/about-citation-files}.

\bibitem[Zenodo(2026)]{zenodoGithub}
Zenodo (2026).
\newblock {Enable a repository}.
\newblock Official documentation on GitHub integration.
\newblock \url{https://help.zenodo.org/docs/github/enable-repository/}.

\end{thebibliography}
\end{document}